# Developing Reprogrammable Metasurfaces with Compressed Pre-curved Beams: Theory and Applications

Fan Liu, Zian Jia, Xihang Jiang, and Lifeng Wang*

Department of Mechanical Engineering, Stony Brook University, Stony Brook, NY 11794, USA

* Corresponding authors. E-mail address lifeng.wang@stonybrook.edu

**Abstract:**

This paper presents a mechanically bistable mechanism of the compressed pre-curved beam. A governing equation is proposed which can be used to predict and explain the bistability of the compressed pre-curved beam. FE simulations and experimental tests are performed to validate the analytical solution. The beam's unique tunable and asymmetrical potential energy landscape is demonstrated which enables the compressed pre-curved beam to switch its stable position without the need of applying an external force to it. Based on that, a coupled beam element is designed and used as a building block to develop reprogrammable metasurfaces that can be programmed to exhibit different configurations.

## 1. Introduction

Bistability refers to the property of a system that has two equilibrium states. It is the essential characteristic of various physical systems such as semiconductor memory devices [1], hybrid bistable optical devices [2], and vibrational energy harvesting devices [3]. In bistable mechanical systems, the two stable equilibrium states usually correspond to two local minima of potential energy and two geometrically different configurations. Complex bistable or multistable structures usually consist of basic bistable elements such as constrained 1D beams, curved 2D plates, and 3D dome shells. Great effort has been put into understanding the bistable mechanisms of these fundamental units and unrevealing the geometrical design principles. Starting from the simplest compressed buckled beam [Fig. 1(a)], various types of beams with different structural forms and complex bistable mechanisms have been designed and investigated. For instance, the coupled beam has been designed to accommodate a linear movement and prevent asymmetric modes from developing [4, 5]. A tilted curved beam has been designed that enables a secondary equilibrium state that is more stable than the primary, stress-free, configuration[6].

Based on the basic bistable elements and well-studied design principles, various bistable structures have been widely exploited for a variety of applications owing to their distinct properties from linear structures. First, when the stable equilibrium state switches from one to the other, large deformation usually occurs. Also, such deformation or shape change does not need to consume additional energy to maintain [Fig. 1(b)]. Owing to that, the bistable structures have been extensively used for microelectromechanical systems [7-9]. Second, some types of bistable structures, such as pre-curved beams, have asymmetrical energy landscapes. During the snapping from a lower energy state to a higher energy state by an external force, the energy difference $\Delta U$ is stored in the structure [Fig. 1(c)]. The energy trapping mechanism has inspired the design of many energy absorption applications [10-12]. Third, the snap-through instabilities of the bistable structures enable the structure to release the stored strain energy within a fast timescale of milliseconds which can generate a large force [Fig.1(d)]. Based on such snap-through instabilities, a variety of soft robotics have been developed [13-15].

Though the bistable mechanisms of compressed beams and pre-curved beams have been well studied, their combination, the compressed pre-curved beam, has not attracted much attention. In this letter, theoretical arguments, numerical simulations, and experimental tests are performed to provide a comprehensive study of the bistability of the compressed pre-curved beam. The combined analytical, numerical, and experimental results indicate that the compressed pre-curved beam with a certain as-fabricated shape has a distinct tunable asymmetrical energy landscape, which enables such structures to switch their energy state and cross the energy barrier without an external force. Moreover, the pre-curved beam is used as a building block to design 2-D programmable metasurfaces that have post-fabrication programmability.

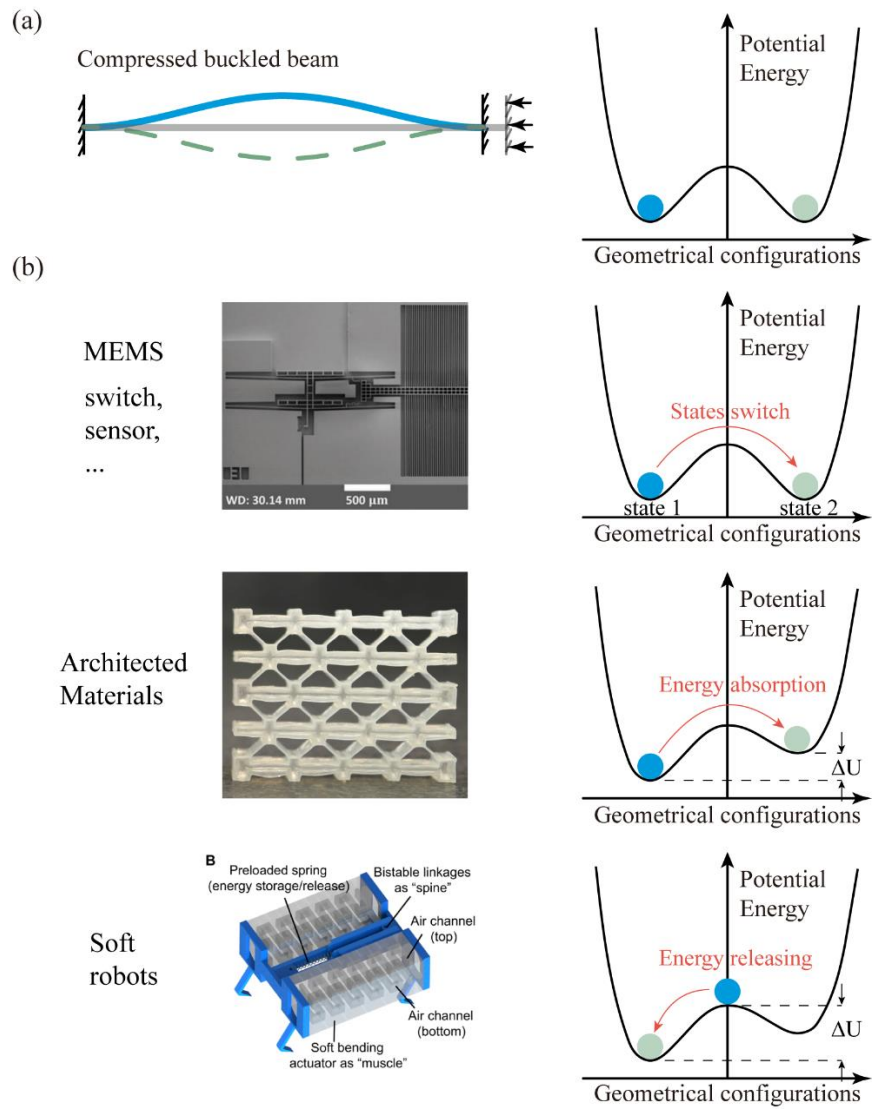

Figure 1. (a) Compressed buckled beam and its potential energy landscape. (b) Applications of bistable structures: microelectromechanical systems[9], architected material for energy absorption[12], and soft robots[15].

## 2. Theoretical Modeling

Before studying the bistability of the compressed pre-curved beam, two simple types of bistable beams are studied: the double-clamped straight beam [Fig. 2(c)] and the pre-curved beam [Fig. 2(d)]. For the double-clamped beam, when the critical load is reached, the straight beam fails by buckling and turns into a bistable beam. The critical load is [16]:

$$P_{cr} = \pi^2 \frac{4EI}{l^2} \qquad (1)$$

where $E$ is Young's modulus of the beam material, $I$ is the moment of inertia of the beam, and $l$ is the initial length of the beam.

For a pre-curved beam (with the second mode constrained), it is bistable if [4]:

$$\frac{h}{t} \geq \frac{4}{\sqrt{3}} \qquad (2)$$

where $h$ is the initial apex height of the beam and $t$ is the constant thickness of the beam.

Although the bistable mechanisms of these two types of beams have been thoroughly studied, their potential energy landscapes haven't been directly reported. The normalized potential energy of the double-clamped beam is:

$$U_c = \frac{4}{l^2}\left(\frac{w_m}{2}\right)^2 \pi^4 + \frac{6}{t^2 l^2}\left(dl - \left(\frac{w_m}{2}\right)^2 \pi^2\right)^2 \qquad (3)$$

where $U_c$ is the normalized potential energy, $U_c = u_c \dfrac{2l}{EI}$ ($u_c$ is potential energy); $w_m$ is the position of the middle point of the beam; $d$ is the compression of the beam; $t$ is the thickness of the beam. The bistability of the double-clamped beam is controlled by compression $d$. The potential energy landscapes of the double-clamped beam with different initial compression $d$ are shown in Fig. 3(a). With the increase of $d$, the shape of the potential energy landscape changes from a 'U' shape to a 'W' shape. The two local minima of potential energy indicate the two stable equilibrium states. Also, $|\partial U_c/\partial w_m|$ is calculated and shown in Fig. 3(b), while $|\partial U_c/\partial w_m|=0$ indicating an equilibrium state. With the increase of $d$, the number of equilibrium states changes from 1 to 3, and the changing point is just located at the critical displacement derived from Eq. (1). This point is the boundary of non-bistable compressed beams and bistable compressed beams.

The normalized potential energy of the pre-curved beam is:

$$U_p = \frac{4\pi^4}{l^2}\left(\frac{h}{2}-\frac{w_m}{2}\right)^2 + \frac{6}{t^2 l^2}\cdot\left(\frac{\pi^2 h^2}{4}-\left(\frac{w_m}{2}\right)^2 \pi^2\right)^2 \tag{4}$$

Here $h/t$ controls the bistability of the pre-curved beam [4]. As shown in Fig. 2(e-f), the number of equilibrium states changes from 1 to 3 when $h/t$ exceeds the value of 2.31 from Eq. (2).

These two types of beams have their unique bistabilities respectively. The double-clamped beam has post-fabrication tunable bistability, by changing the compression $d$, the potential energy landscape can be changed and the bistability of the beam can switch between bistable and non-bistable. The pre-curve beam's bistability is not tunable, however, its potential energy landscape is asymmetrical: the two stable equilibrium states have different potential energy. Having investigated the bistability of these two types of beams, the bistability of the compressed pre-curved beam [Fig. 4(a)] is explored and the combined effect of $l/t$, $d/l$, and $h/t$ on the bistability is studied.

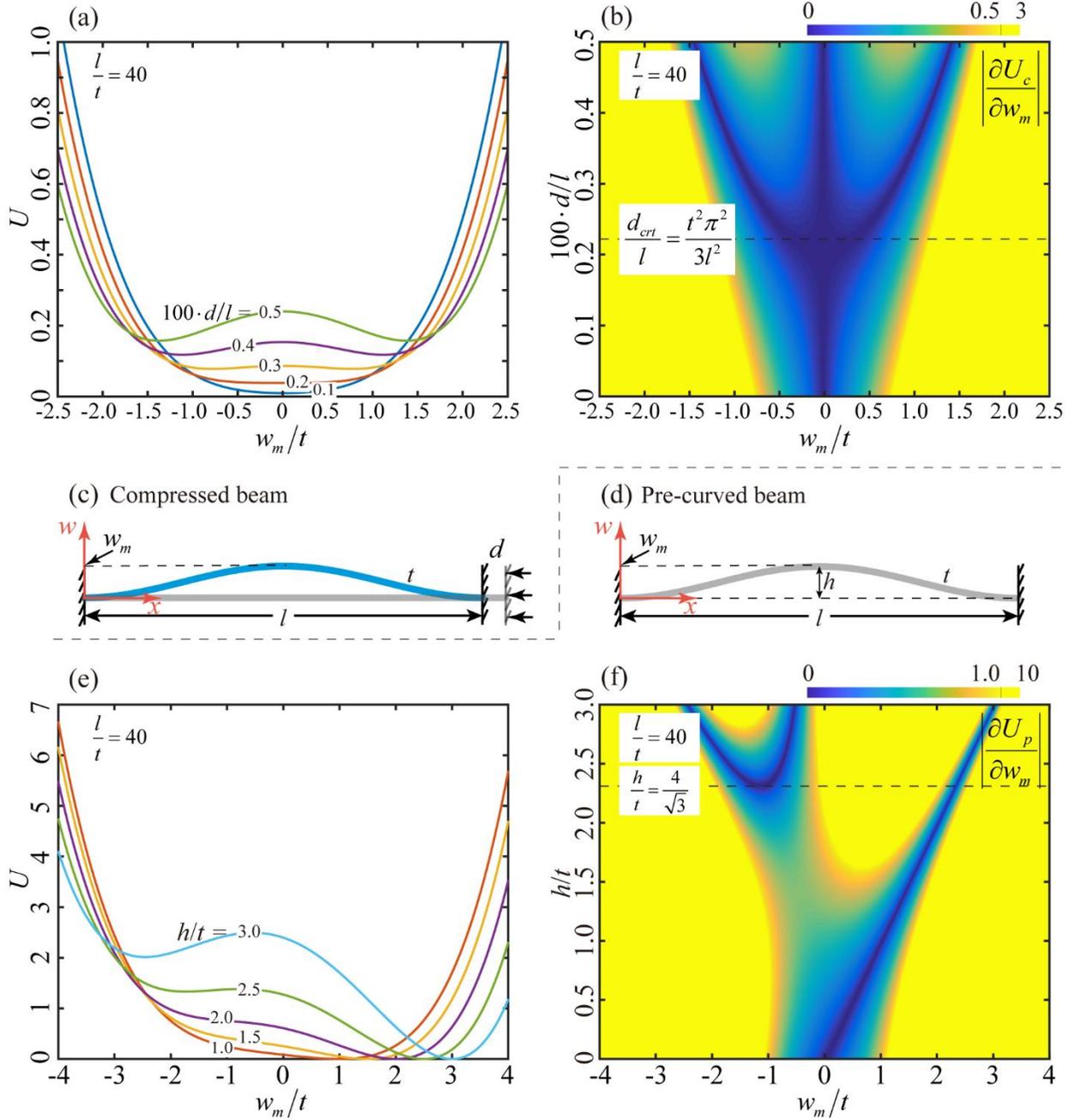

Figure 2. Potential energy landscapes of the compressed beam and the pre-curved beam. (a-b) Potential energy landscapes and $\left|\partial U_c/\partial w_m\right|$ of the compressed beam with different compression $d$. (c-d) Schematic diagrams of the compressed beam and the pre-curved beam. (e-f) Potential energy landscapes and $\left|\partial U_p/\partial w_m\right|$ of the pre-curved beam with different $h/t$.

The equilibrium equation describing a beam subjected to an axial load is:

$$EI\frac{d^4w}{dx^4} + p\frac{d^2w}{dx^2} = 0 \tag{5}$$

where $p$ is the axial load and $w$ is the deflection. The boundary conditions of the beam as shown in Fig. 3(a) are $w(0) = w(l) = w'(0) = w'(l) = 0$. The non-zero solutions can be divided into two groups:

$$\omega_j = \begin{cases} A_j\left(1-\cos\left(N_j\frac{x}{l}\right)\right) & N_j = (j+1)\pi & j=1,3,5,7... \\ \left(1-\cos\left(N_j\frac{x}{l}\right) + \frac{2}{N_j}\sin\left(N_j\frac{x}{l}\right) - 2\frac{x}{l}\right) & N_j = 2.86\pi, 4.92\pi, 6.94\pi... & j=2,4,6,8... \end{cases} \tag{6}$$

The shape of the beam is $\omega(x) = \sum_{j=1}^{\infty} \omega_j$. Here, the beam is pre-curved, and the as-fabricated shape is:

$$\bar{\omega} = \frac{h}{2}\left(1-\cos\left(\frac{2\pi x}{l}\right)\right) \tag{7}$$

After the compression, the total length of the beam changes to:

$$l + \bar{d} + d - d_p = \int_0^l \left\{1 + \frac{[\omega'(x)]^2}{2}\right\}dx = l + \sum_{j=1}^{\infty}\frac{(A_j N_j)^2}{4l} \tag{8}$$

where, $l + \bar{d}$ is the initial length of the beam before compression:

$$l + \bar{d} = \int_0^l \left\{1 + \frac{[\bar{\omega}'(x)]^2}{2}\right\}dx = l + \frac{\pi^2 h^2}{4l} \tag{9}$$

$d_p$ is the length change of the beam due to the axial force:

$$d_p = d + \frac{\pi^2 h^2}{4l} - \sum_{j=1}^{\infty} \frac{(A_j N_j)^2}{4l} \tag{10}$$

The potential energy of the compressed pre-curved beam can be written as:

$$u_{cp} = \frac{EI}{2} \int_0^l \left[\omega''(x) - \bar{\omega}''(x)\right]^2 dx + \frac{1}{2} p \cdot d_p \tag{11}$$

where the first term is bending energy and the second term is compression energy. The axial force is $p = Ebtd_p/l$. From Eq. (6), Eq. (10), and Eq. (11), and considering there is no lateral load, the normalized potential energy can now be expressed as

$$U_{cp} = \frac{4\pi^4}{l^2}\left(\frac{h}{2} - \frac{w_m}{2}\right)^2 + \frac{6}{t^2 l^2}\left(dl + \frac{\pi^2 h^2}{4} - \left(\frac{w_m}{2}\right)^2 \pi^2\right)^2 \tag{12}$$

Mathematically, the beam is bistable when the equation $|\partial U/\partial w_m| = 0$ has more than 1 distinct real root, which can be written as

$$\left(\frac{w_m}{2}\right)^3 - \left(\frac{w_m}{2}\right)\left(\frac{dl}{\pi^2} + \frac{h^2}{4} - \frac{1}{3}t^2\right) - \frac{1}{6}t^2 h = 0 \tag{13}$$

this cubic polynomial has three distinct real roots if

$$\frac{d}{l} > \frac{\pi^2}{l^2}\left(\sqrt[3]{\frac{3t^4 h^2}{16}} - \frac{h^2}{4} + \frac{t^2}{3}\right) \tag{14}$$

This is the governing equation of the bistability of the compressed pre-curved beam. By plotting the equation in a 3D space, it has a shape of a dome-shaped surface as shown in Fig. 4(a). For any compressed pre-curved beam, its bistability can be determined by calculating three normalized parameters: $d/l$, $h/t$, and $l/t$. The beam is bistable if the point ($l/t$, $h/t$, $d/l$) is above the surface, on the contrary, the beam is non-bistable if the point ($l/t$, $h/t$, $d/l$) is below the surface. Five cross-sections ($l/t = 20 \sim 60$) of the dome-shaped surface are plotted in Fig. 4(b) to further explain how the three normalized parameters affect the bistability of the compressed pre-curved beam. First, beams with larger $l/t$ have larger bistable regions. Second, for straight beams ($h/t$ =0), the critical compression to achieve bistability is controlled by $l/t$. Third, any beam, with $h/t$ larger than $4/\sqrt{3}$, is bistable. The dome-shaped surface is defined as the boundary surface of the bistability and its cross-sections are defined as the boundary curves of the bistability.

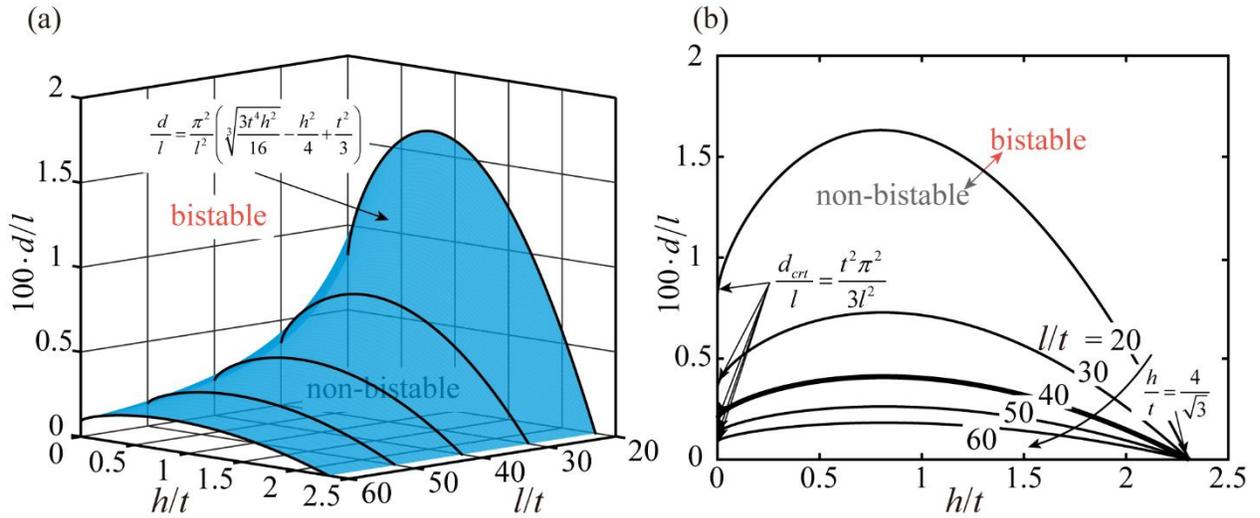

Figure 3. (a) The governing equation of the bistability of the compressed pre-curved beam. (b) Five cross-sections ($l/t = 20 \sim 60$) of the dome-shaped surface. The bistable region is above the curve and the non-bistable region is below the curve.

To verify the governing equation, the $\left|\partial U_{cp}/\partial w_m\right|$ of beams with $l/t = 40$, $h/t = 0.5 \sim 2.5$, and $100 \cdot d/l = 0 \sim 0.6$ are calculated and shown in Fig. 4(b-f). Obviously, these results match well with the boundary curve ($l/t = 40$) as shown in Fig. 3(b). For a beam with a small $h/t$, the compression $d$ must be large enough to enable the bistability of the beam. For example, as shown in Fig. 3(b), when $h/t = 1$, the boundary of the bistability is located at $100 \cdot d/l \approx 0.4$. Correspondingly, as shown in Fig. 4(c), the beam gains a second stable equilibrium state when the compression reaches $100 \cdot d/l \approx 0.4$. However, for a beam with a larger $h/t$ ($h/t = 2.5$), the beam is bistable without compression [Fig. 3(b) and Fig. 4(f)].

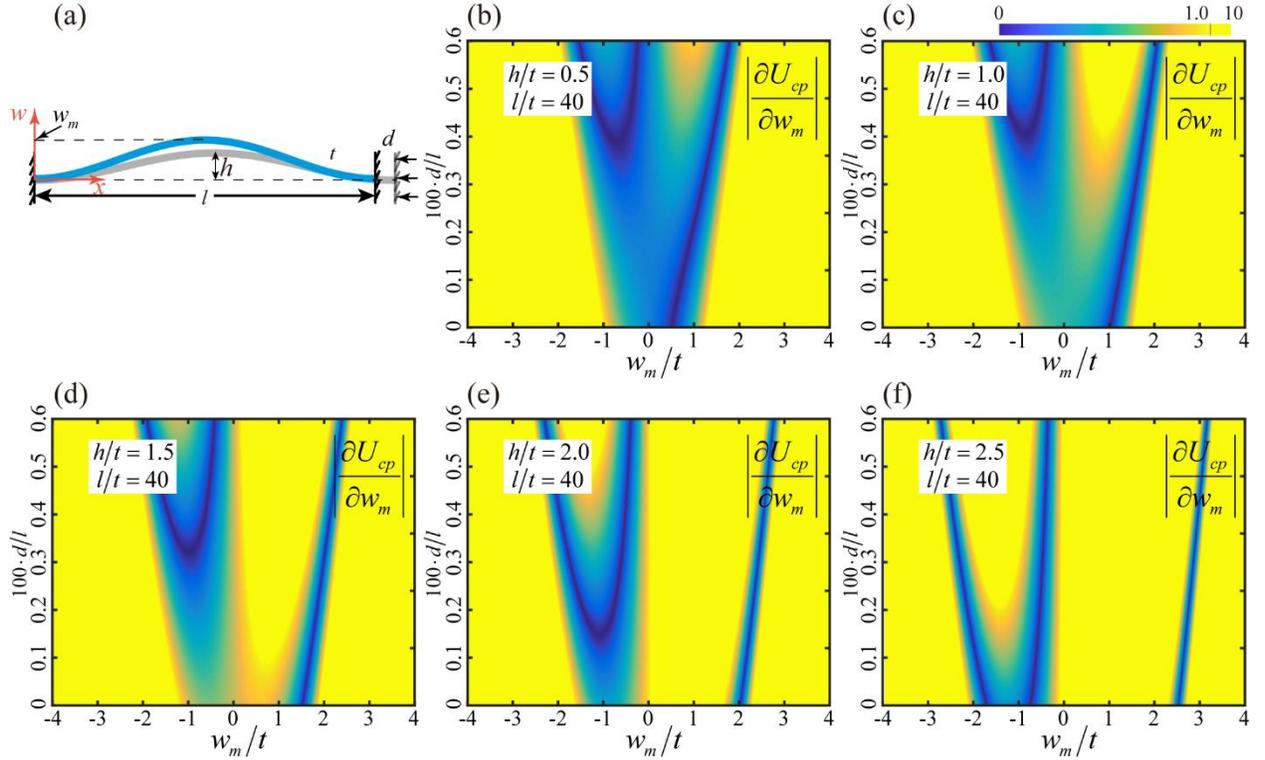

Figure 4. (a) Schematic diagrams of the compressed pre-curved beam. (b) $|\partial U/\partial w_m|$ of the beams with different $d_0$ and $h/t$.

## 3. FE simulations and experimental tests

To further verify the analytical solution, a nonlinear finite element analysis is performed using ABAQUS. The FE beam model (with Timoshenko beam element B21) has the following geometrical and material properties: $l = 40mm$, $b = 5mm$, $t = 1mm$, $E = 2000MPa$, and $v = 0.3$.

A lateral displacement is applied at the center of the beam and the force-displacement curves are obtained from the simulation results. A group of normalized force-displacement curves of beams with $h/t = 0.5$ and $100 \cdot d/l = 0 \sim 0.6$ are shown in Fig. 5(c). The first 20 curves (gray color) with $100 \cdot d/l < 0.4$ are always above 0 which means these beams are non-bistable. The 11 curves with $100 \cdot d/l \geq 0.4$ (red color) have large portions below zero force indicating these two beams are bistable. This is how to determine if the beam is bistable from the simulation results. Based on this criterion, a series of simulations (279 simulations in total) are conducted with $h/t$ in the range of 0.25 to 2.25 and $100 \cdot d/l$ in the range of 0 to 0.6 to determine the bistability of the beams. All the simulation results are shown in Fig. 5(b), where bistable beams are marked with red dots and non-bistable beams are marked with gray dots. The boundary curve from Fig. 3(b) perfectly lies on the boundary between the red and gray dots and separates the region into an upper bistable one and a lower non-bistable one. The analytical results match well with the simulation results.

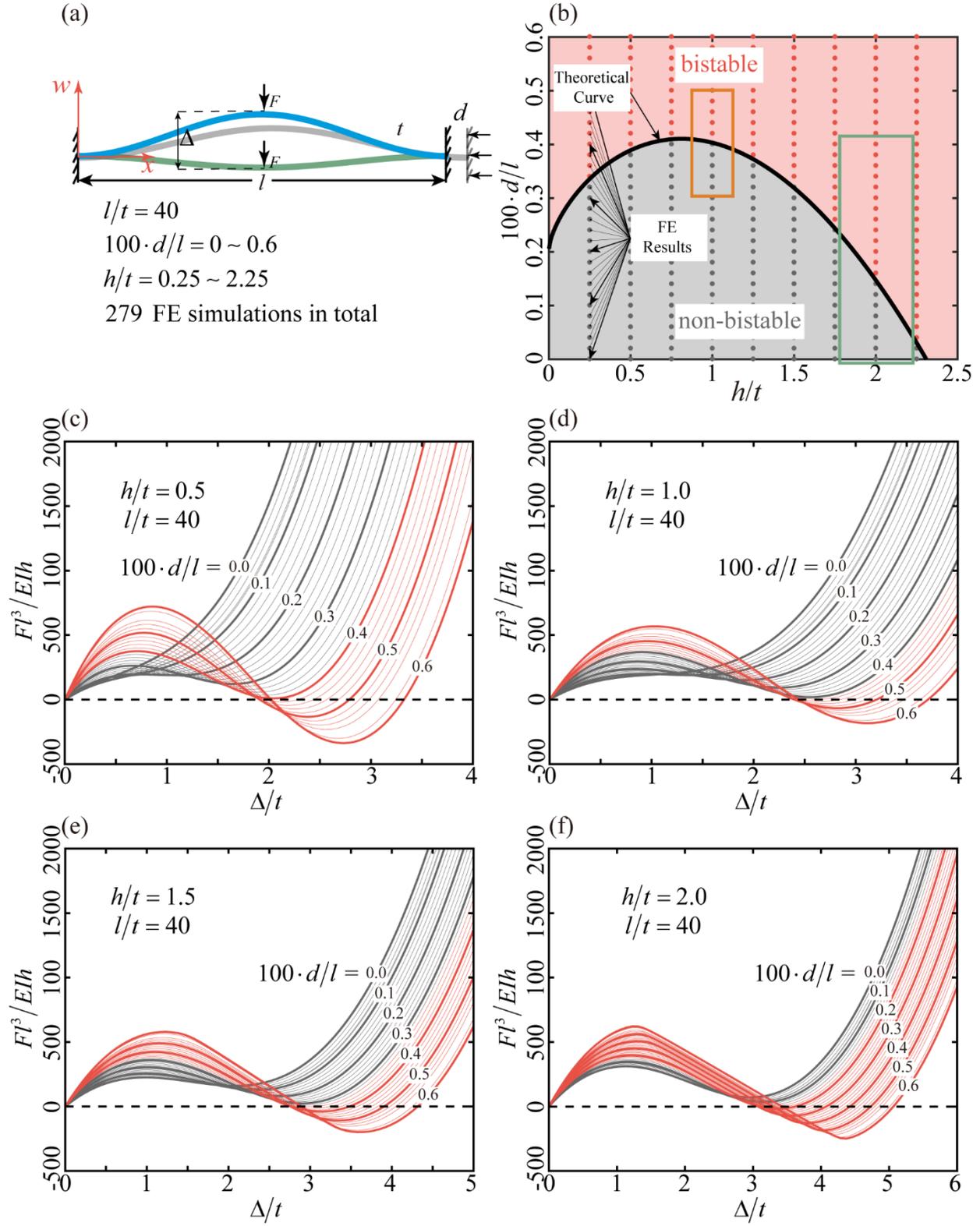

Figure 5. (a) Schematic diagrams of the geometry, boundary conditions, and loading of the FE model to determine the bistability of the compressed pre-curved beam. (b) FE simulations results and the theoretical boundary curve of the beams with $l/t = 40$. (c-f) Normalized force-displacement curves of compressed pre-curved beams with lateral forces applied on.

Experimental tests are also performed to validate the analytical solution. The specimen used in this study were fabricated using an Anycubic Photon M3 Plus 3D printer which has a resolution of 34 microns in the XY direction and 10 microns in the Z direction. The fabricated 3D printed beam has the following geometrical and material properties: $l = 100mm$, $h = 5mm$, $t = 2.5mm$, $E = 800MPa$, and $v = 0.3$. The printed specimen has an adjustable boundary for tuning the compression $d$ of the beam. Five tests were performed on the beam with different compression $100 \cdot d/l = 0 \sim 0.4$ and the normalized force-displacement curves from the tests are shown in Fig. 6(b). Beams with compression of $100 \cdot d/l = 0$ and 0.1 are non-bistable while beams with compression of $100 \cdot d/l = 0.2$, 0.3, and 0.4 are bistable. The beams in the tests have $l/t = 40$ and $h/t = 2$ which are located in the region with green boundaries as shown in Fig. 5(b) and Fig. 6(c). The test results and theoretical boundary curve as shown in Fig. 6(c) indicates that the analytical results match well with the experimental results.

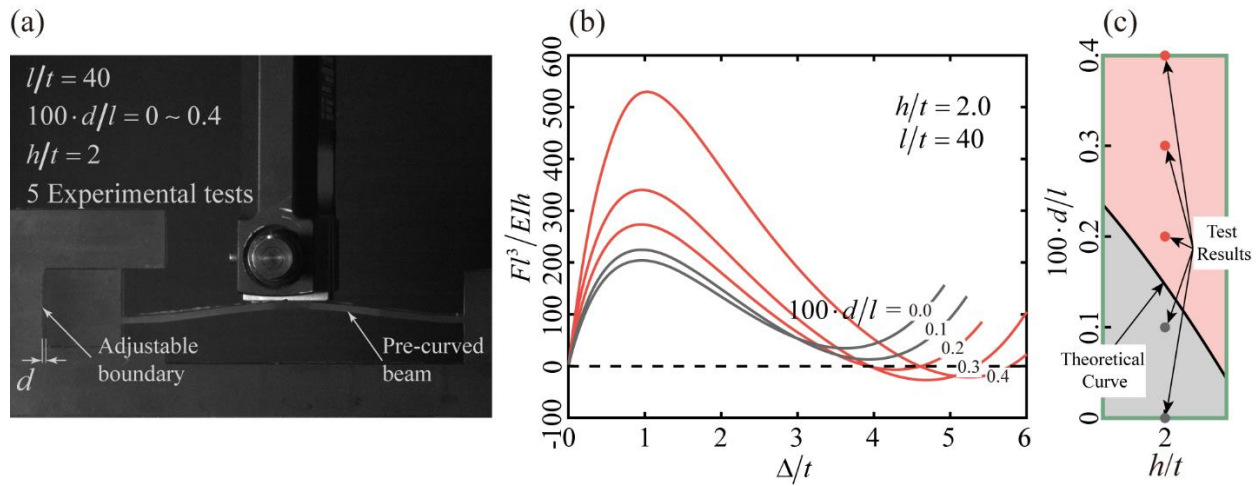

Figure 6. (a) 3D compressed pre-curved beam specimen and experiment setup. (b) Normalized force-displacement curves from the experimental tests. (c) Experimental results and the theoretical boundary curve of the beams with $l/t = 40$ and $h/t = 2$.

Now focus on the region with orange boundaries that is shown in Fig. 5(b) as well as Fig. 7(a). This region indicates a pre-curved beam with $h/t = 1$ and can be compressed between $100 \cdot d/l = 0.3$ to 0.5. When the compression $100 \cdot d/l = 0.45$ (C0), the compressed beam is located in the bistable region and has two stable equilibrium states. The primary stable state, marked with blue color, has lower potential energy, and the secondary state, marked with green color, has higher potential energy. Once the beam is in the secondary state, it's still possible to switch it to the primary state. The simplest way is to apply a lateral force and push the beam upwards to the primary state as shown in Fig. 7(b). A FE simulation is performed to study this state switch process and the results are shown in Fig. 7(b). The curve indicates the force-displacement relationship during the state switch and the configurations of the beam at two stable states. One way to switch the stable state of the beam is by applying an external force that is large enough to overcome the force barrier.

Other than applying an external force, there is another way to switch the secondary state to the primary state without applying external force. To achieve the state switch, the compression of the beam needs to be changed following the process as shown in Fig. 7(a). Simply by decompressing the beam to the compression $100 \cdot d/l = 0.35$ ( C2), which is located in the non-bistable region, and then compressing the beam back to the initial C0, the state of the beam can be switched from the secondary state to the primary state. The change of $w_m$ [Fig. 7(c)] and the change in the potential energy landscape [Fig. 7(d-g)] during the state switch process are obtained from the FE simulations and analytical calculation.

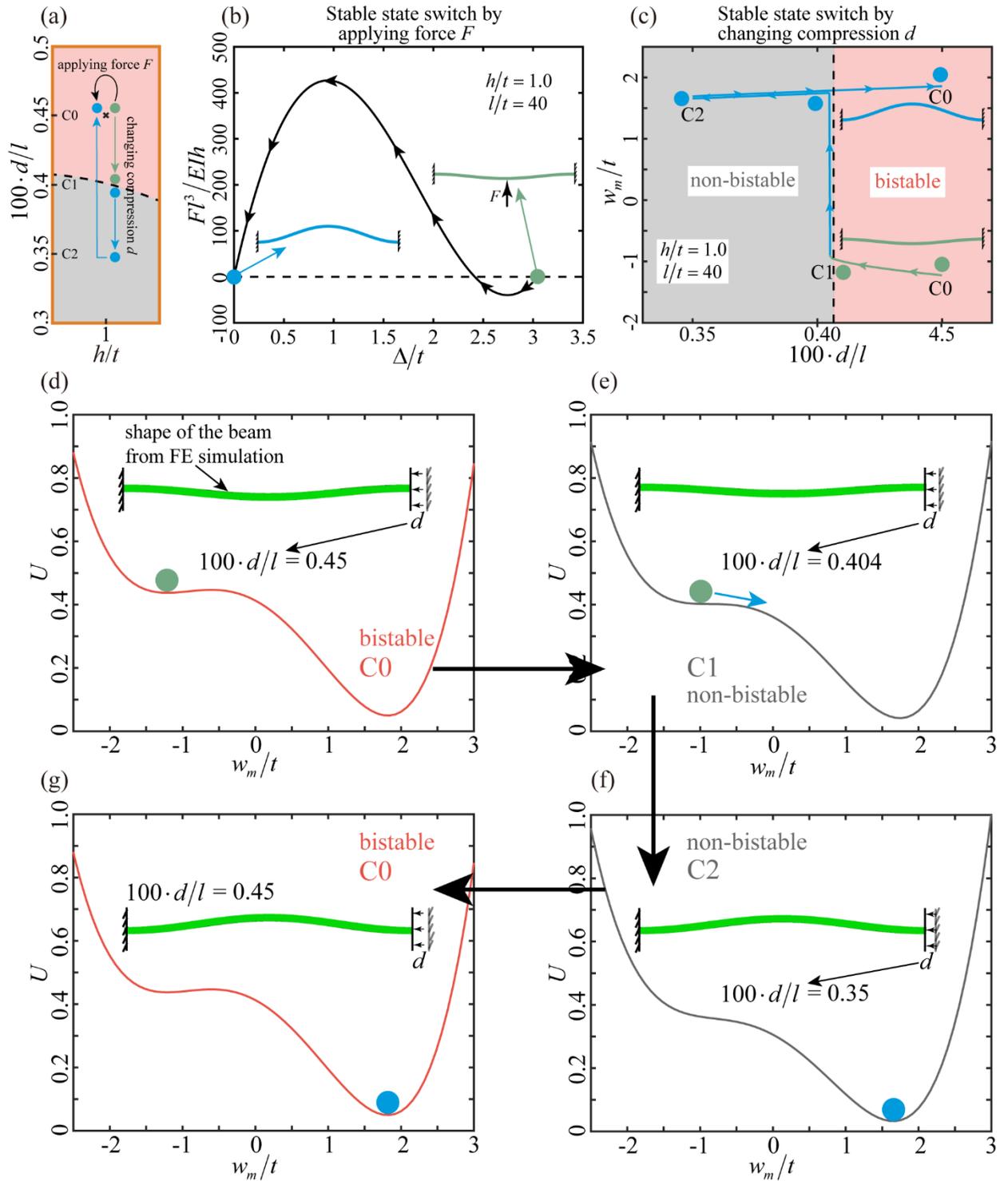

Figure 7. Stable state switch of the compressed pre-curved beam. (a) Schematic diagram of two types of state switch: stable state switch by applying external force and stable state switch by changing compression $d$. (b) Force-displacement curve and beam configurations during the state

switch by applying an external force on the beam. (c) The change of the location of the center of the beam $w_m$ during the state switch by changing the compression $d$ of the beam. (d-g) The change of potential energy landscape and beam configurations during the stable state switch process.

Initially, the beam is in the bistable region at C0, while the position of the center of the beam is $w_m/t \approx -1.2$. The potential energy landscape of the beam with C0 is shown in Fig. 7(d) and the beam is at the secondary stable position with higher potential energy. Then the beam is decompressed, and the center of the beam moves upwards until it reaches C1, the boundary of the bistable and non-bistable region. Then snap-through motion happens when crossing the boundary, and the center of the beam moves upward dramatically. The potential energy landscape of the beam with C1 is shown in Fig. 7(e), the energy barrier between the primary and secondary stable states disappears, and the state of the beam switches from the secondary to the primary state. The quickly released potential energy triggers the snap-through. After that, the beam is decompressed to C2 with only one stable state [Fig. 7(f)]. Lastly, the beam is compressed back to the initial C0, but the beam does not go back to its initial configuration ($w_m/t \approx -1.2$) and the center of the beam stays in $w_m/t \approx 1.9$. The potential energy landscape resumes its initial shape at C0, but the state of the beam stays in the primary state as shown in Fig. 7(g).

## 4. Developing Reprogrammable Metasurface

The tunable and asymmetrical energy landscape of the compressed pre-curved beam enables the energy state switch without an external force. Based on this distinct property, a coupled beam is proposed [Fig. 8(a)] which can be used as a unit cell to construct a 2D programmable metasurface. The bistability of the coupled beam after fabrication (with fixed initial shape $h/t$ and $l/t$) is affected by two parameters: $d_x/l$, and $d_y/l$ (compression in both $x$ and $y$ directions). A compressed pre-curved beam with $h/t=1$ and $l/t=40$ has the boundary of bistability at $100 \cdot d/l = 0.404$ according to Eq. (14). Therefore, the two intersecting lines [dashed lines with blue color as shown in Fig. 8(b)] $100 \cdot d_x/l = 0.404$ and $100 \cdot d_y/l = 0.404$ divide the plane into four sections. In the top-right section, both beams in the $x$ and $y$ directions are bistable which makes the coupled beam element bistable. In the bottom-left section, both beams are non-bistable which makes the element non-bistable. However, in the top-left and bottom-right sections, one beam is bistable and the other one is non-bistable. Therefore, FE simulations are performed to determine the bistability of the couple beam elements in these two sections. The simulation results are marked with red and gray dots in Fig. 8(b) which outlines the boundary of bistability (dashed lines with black color).

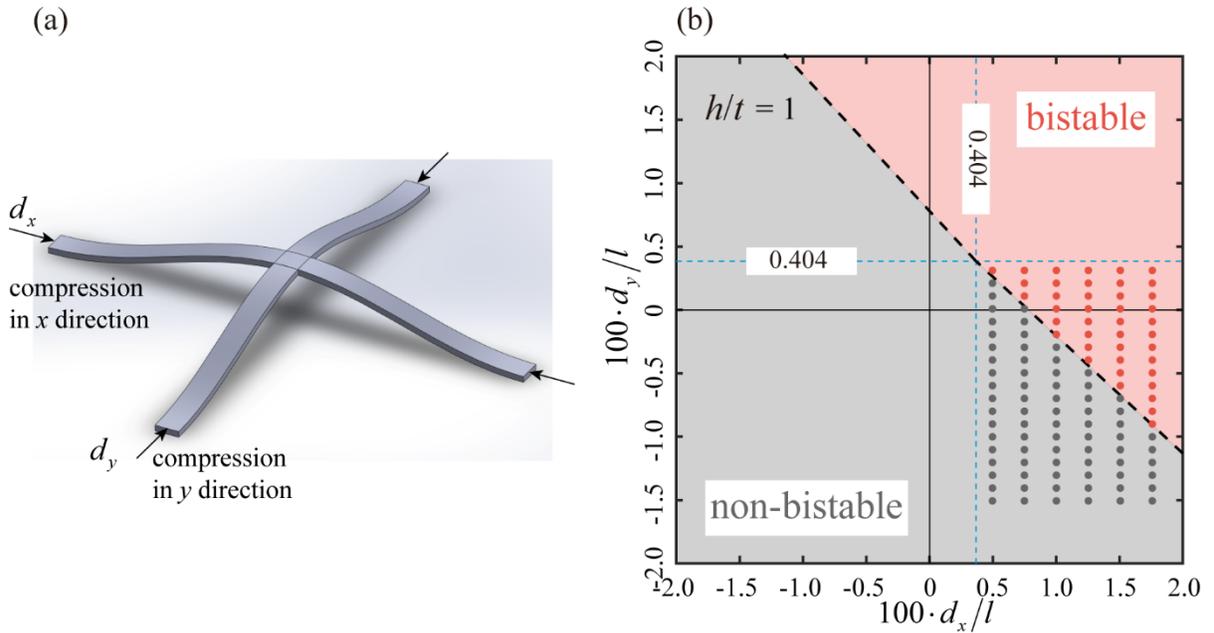

Figure 8. (a) Schematic diagrams of the coupled beam element. (b) Effect of $d_x/l$, and $d_y/l$ on the bistability of the coupled beam element with $h/t=1$ and $l/t=40$.

Fig. 9(a) shows the simplest meta-surface assembled by 4 of the coupled beam elements. A series of designed compressions are applied to the structure. Initially, all 4 elements of the 2×2 metasurface are in the secondary stable states with the compression of C0. Then certain displacements are applied to the boundaries and the compression of the structure changes to C1. Note that, the stable state of Element A switches from secondary to primary state at C1, but the other three elements stay in secondary stable state. Lastly, the compression of the structure are resumed to the initial C0, however, the stable state of Element A stays in the primary state. By tracking the compressions of the four elements in *x* and *y* directions ($d_x/l$ and $d_y/l$), the reason why Element A switched its stable state, but not the other three elements can be answered. As shown in Fig. 9(b), Element A is the only one that crosses the boundary of the bistability and therefore changes the stable state. For a 2×2 metasurface, each element has two stable positions, therefore, there are in total of 16 configurations. By applying different compression to the structure, different configurations can be achieved as shown in Fig. 9(c). In other words, the configuration of the structure is programmable.

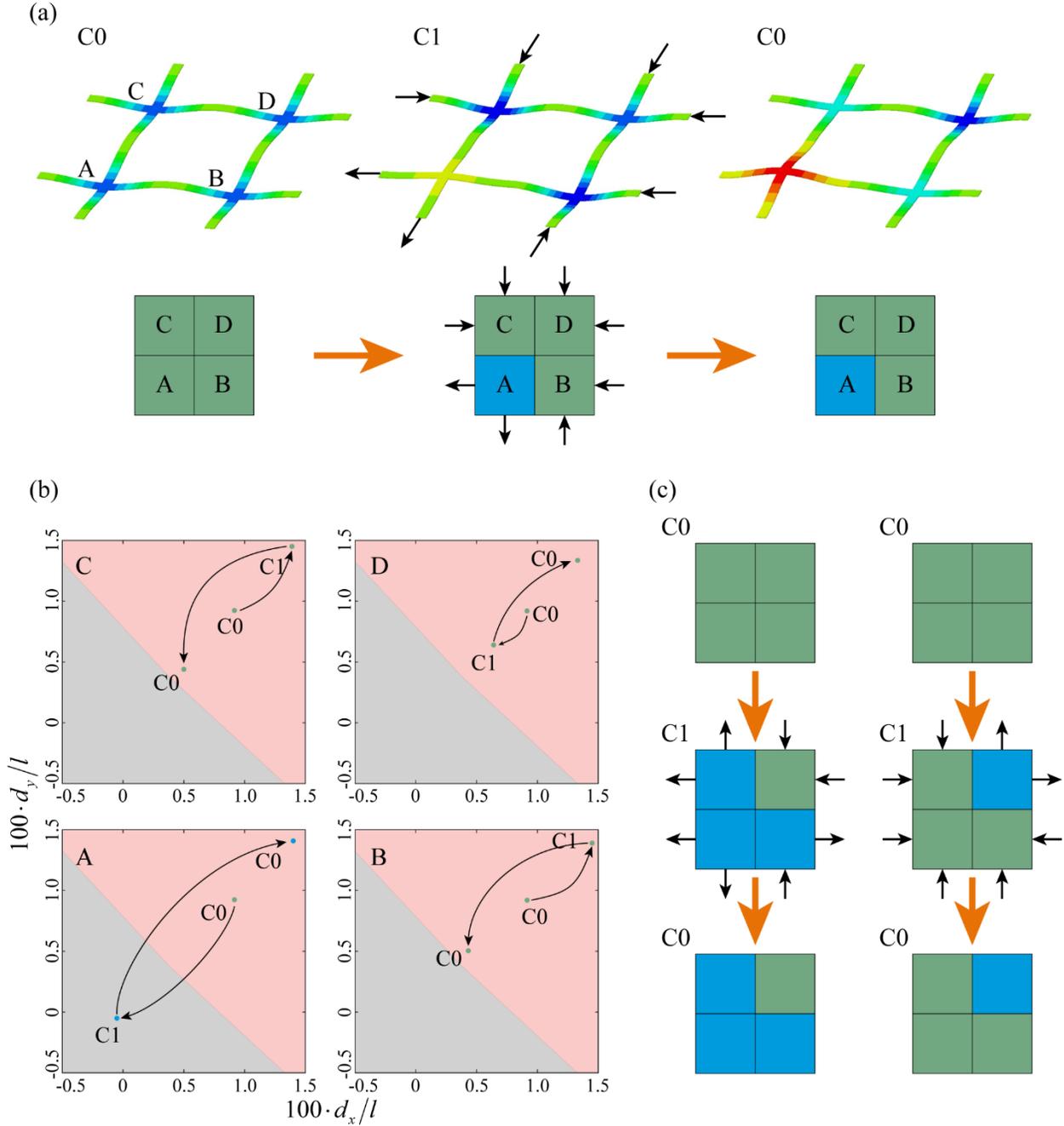

Figure 9. (a) A 2×2 metasurface made of coupled beams that can be programmed by intermediate compressions. (b) Stable state switch of each of the four elements during the compression changing process. (c) Different intermediate compressions lead to different final configurations.

With the metasurface expanding to 5×5, all possible configurations are $2^{25}$ which is a huge design space to be exploited. Two different FE simulations are performed on the 5×5 structures as shown in Fig. 10(a). By applying well-designed sequences of compressions on the structure, configurations with the shape of numbers '1' and '2' are achieved respectively. For a larger structure, designing the compression sequences for specific configurations could be complex and very time-consuming. However, a much easier way to switch the stable state of a specific element in the structure is by creating a local expansion. Fig. 10(b) shows the FE simulation on an 8×9 structure. By applying local expansion to a series of elements one by one, a configuration with a shape of a smiley face is created. In practice, such local expansion can be applied to a real structure by using a laser beam to heat the specific local areas.

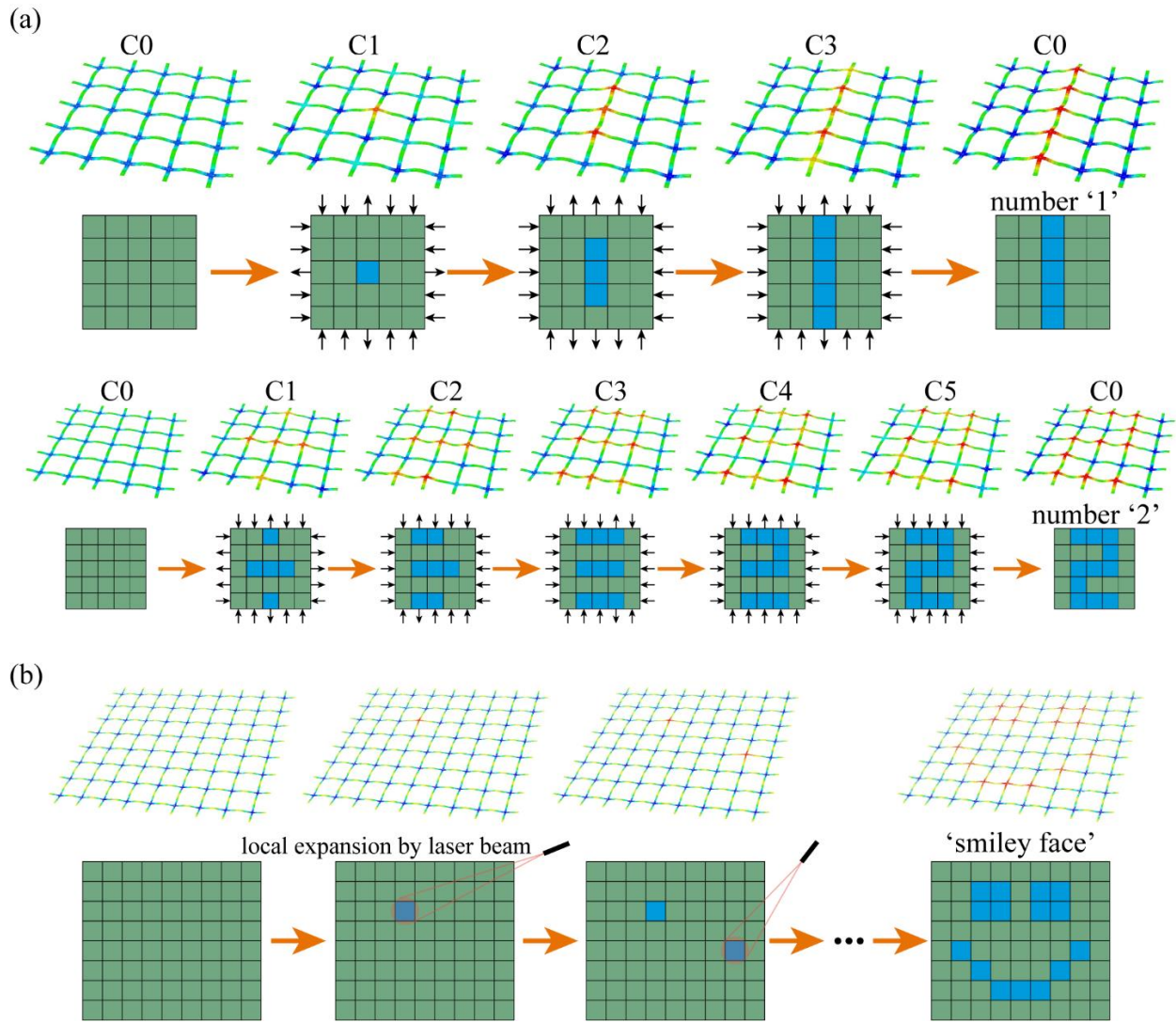

Figure 10. Larger metasurfaces with more complex configurations. (a) By applying different sequences of compressions on a 5×5 metasurface, configurations with the shape of numbers '1' and '2' are achieved respectively. (b) By applying local expansion to a series of elements one by one, a configuration with a shape of a smiley face is created

## 5. Conclusion

In this study, a new concept of the compressed pre-curved beam is proposed. A combined analytical, numerical, and experimental study indicates that three normalized parameters $l/t$, $h/t$, and $d/l$ affect the bistability of the beam. The special bistable mechanism originated from the beam's unique tunable and asymmetrical potential energy landscape is demonstrated. This bistable mechanism enables the compressed pre-curved beam to switch its stable position without the need of applying an external force to it. Further, a coupled beam element is designed and its bistable mechanism is investigated. In the end, the beam is used as a building block to create a metasurface that can be programmed to exhibit different configurations with different intermediate compressions applied to it. The programmable textured surface can be used as a biomaterial to adapt to the complex and changing human body. Also, it could be used at smaller scales to control friction and wetting.